\documentclass[useAMS,usenatbib]{mnras}
\usepackage{graphicx}	
\usepackage{multicol}	
\usepackage{natbib}
\usepackage[english]{babel}
\usepackage{epstopdf}
\usepackage[T1]{fontenc}
\usepackage{ae,aecompl}
\usepackage{amsmath}    
\usepackage{amssymb}    

\newcommand{\rmag}{\>^{0.1}{\rm M}_r-5\log h}

\usepackage{xcolor}

\makeatletter

\newcommand{\Rmnum}[1]{\expandafter\@slowromancap\romannumeral #1@}
\makeatother

\title{Populating HI gas in dark matter halos: I. method}

\author[Lu et al.]{\parbox{\textwidth}{Yi Lu$^{1}$\thanks{luyi@shao.ac.cn},
Xiaohu Yang$^{2,3}$\thanks{xyang@sjtu.edu.cn},
Chengze Liu$^{2}$,
Hong Guo$^{1}$,
Haojie Xu$^{2}$,
Antonios Katsianis$^{3,2}$,
Zhaoyu Wang$^{2}$}
\vspace{0.4cm}\\
\parbox{\textwidth}{$^{1}$Key Laboratory for Research in Galaxies and Cosmology,
  Shanghai Astronomical Observatory; Nandan Road 80, Shanghai 200030,
  China \\
$^{2}$Department of Astronomy, School of Physics and
  Astronomy, Shanghai Jiao Tong University, Shanghai 200240, China \\
$^{3}$Tsung-Dao Lee Institute, and Shanghai Key Laboratory
  for Particle Physics and Cosmology, Shanghai Jiao Tong University,  Shanghai 200240, China}
}

\begin{document}
\label{firstpage}
\pagerange{\pageref{firstpage}--\pageref{lastpage}}
\maketitle

\begin{abstract}
We combine data from the Sloan Digital Sky Survey (SDSS) and the Arecibo Legacy Fast ALFA Survey (ALFALFA) to establish an empirical model for the HI gas content within dark matter halos. A cross-match between our SDSS DR7 galaxy group sample and the ALFALFA HI sources  provides a catalog of 16,520 HI-galaxy pairs within 14,270 galaxy groups (halos).  Using these matched pairs, we model the HI gas mass distributions within halos using two components: 1) {\it in situ} galaxy relations that involve the HI masses, colors $({\rm g-r})$ and stellar masses 2) an {\it ex situ} dependence of the HI mass on the halo mass/environment. We find that if we solely use galaxy associated scaling relations to predict the HI gas distribution (solely component 1), the number of HI detections is significantly over-predicted with respect the ALFALFA observations. We introduce a concept for the survival of the HI masses/members within halos of different masses labelled as the  `efficiency' factor, in order to describe  the probability that a halo has in retaining its HI detections. Taking the above consideration into account we construct a `halo based HI mass model' which does not only predict the HI masses of galaxies, but also yields similar number, stellar, halo mass and satellite fraction distributions to the HI detections retrieved from observational data.
\end{abstract}



\section{Introduction}
\label{sec:intro}

According to the current paradigm of galaxy formation, dark matter halos form and grow through gravitational instabilities from small perturbations (Planck Collaboration et al.2016). Within the potential wells of these halos, gas cools and condenses, while galaxies and stars form (Katsianis et al. 2017; Zhao et al. 2020). Thus, it is worth to investigate any underlying gas-galaxy-halo connections, since they can provide insight about the physical processes that regulate galaxy formation and evolution (Wechsler \& Tinker 2018).

The galaxy-halo connection has been extensively studied in the literature either by employing analytical models such as the halo occupation distribution method (HOD, e.g. Jing et al. 1998; Berlind \& Weinberg 2002; Zheng et al. 2005; Zehavi et al. 2011; Guo et al. 2015), the conditional luminosity function (CLF, e.g. Yang et al. 2003), the subhalo abundance matching (SHAM; Vale \& Ostriker 2004) or direct observational measurements obtained for galaxy groups\footnote{Groups are defined as sets of galaxies that reside within the same dark matter halo (Yang et al. 2006, 2008).}.

Based on large sky redshift surveys such as the Sloan Digital Sky Survey (SDSS; York et al. 2000), numerous group catalogs have been constructed (Berlind et al. 2006; Yang et al. 2007; Crook et al. 2008). Among different group finders that have been applied to large redshift surveys (Eke et al. 2004; Crook et al. 2007; Berlind et al. 2006; Lavaux \& Hudson 2011), the halo-based group finder developed by  (Yang et al. 2005), which sorts galaxies to their host halos, has been extensively tested using mock galaxies from simulations and has been found to perform typically better than other traditional friends-of-friends method. Thus, the group finder described in (Yang et al. 2005) is suitable to study the relation between galaxies and dark matter haloes over a wide dynamic range of halo masses, from rich clusters to groups that contain just a few galaxies. The group finder has been already applied successfully to 2dFGRS (Yang et al. 2005), SDSS DR4 \& DR7 (Yang et al. 2007), 2MRS (Lu et al. 2016) and 6dF (Lim et al. 2017) to construct well-defined galaxy group catalogs. These catalogs can provide useful constraints on the galaxy-halo connection and offer the opportunity to study  galaxy evolution within dark matter halos.

Contrary to the well studied galaxy-halo connection, due to difficulties in gas observations, the gas-galaxy connection is not successfully constrained. Studies that focus on the gas-galaxy connection can only give insights for a certain mass range of halos (usually with a halo mass $M_h \geq 10^{13}\,M_\odot$), e.g., using X-ray observations to probe the hot gas in clusters  (Wang et al. 2011; Wang et al. 2014) or employing CO, HCN observations to probe cold molecular gas in small clumps in galaxies  (Saintonge et al. 2017; Tacconi et al. 2018; Piotrowska et al. 2020), or emission line properties (Lopez et al. 2020). The neutral atomic hydrogen (HI), which is thought to be loosely gravitationally bound to galaxies, can be an important component to construct models for the gas-galaxy connection and involve a much larger halo mass range than other gas elements. In addition, the HI content is related to both the host halo properties and regulates key properties of the galaxies residing within (Guo et al. 2017,2020). For example, when galaxies move towards the center of the gravitational potential of larger halos, the cold gas within galaxies can be easily re-disturbed by galaxy interactions and tidal forces (Hibbard \& van Gorkom 1996). Furthermore, HI gas in a galaxy, if cooled efficiently, can provide the reservoir for star formation (Katsianis et al. 2017a). Therefore, the HI content is a good proxy for a halo's impact to its member galaxies while directly correlates with galaxy properties.

In the past decade, large HI surveys have provided the HI masses for thousands of galaxies with large multi-wavelength coverage, such surveys include the HI Parkes All-Sky Survey (HIPASS, Meyer et al. 2004) which involved $\sim$ 5,000 extra-galactic HI sources out to $z \sim 0.04$ and covers the whole southern sky and the Arecibo Fast Legacy ALFA Survey (Giovanelli et al. 2005) which detected more than 30,000 extra-galactic HI sources out to $z \sim 0.06$ in the northern sky. Based on the above surveys, our understanding for the cold gas content of galaxies has improved. For example, the HI mass function of the HI-rich galaxies in the local universe has been constrained (Zwaan et al. 2005; Martin et al. 2010). In addition, numerous studies have been carried out in order to unravel how the HI content in the galaxies varies with morphology, luminosity, size and star formation activity (Boselli et al. 2001; Cortese et al. 2011; Wang et al. 2015). Thus, some studies have established correlations between the HI and galaxies properties (Kannappan et al. 2004; Zhang et al. 2007;  Zhang et al. 2009; Catinella et al. 2010; Catinella et al. 2012). In addition, other authors focus on establishing relations between the gas content and group/cluster environment (Rasmussen et al. 2012; Serra et al. 2012; Brown et al. 2016; Stark et al. 2016). Through statistical analyses of the HI gas content of member galaxies within clusters like Virgo and Coma, it has been noticed that most massive groups are deficient in HI, especially toward the center (Solanes et al. 2001; Gavazzi et al. 2013; Taylor et al. 2012; Cortese et al. 2008), while the situation is still unclear in smaller halos (Rasmussen et al.2012b).

Instead of studying specific clusters, some authors have examined the impact of environment for a wide range of galaxy group halo masses using statistical samples. Using a control sample which is constructed by isolated field galaxies with similar stellar
masses and redshifts, some studies find that the HI content can be affected by the properties of the host halo/group (Catinella et al. 2010,2012; Hess \& Wilcots 2013; Yoon et al. 2015) or local density (Fabello et al. 2012). Several mechanisms have been suggested to contribute to the HI-deficient within a group, such as ram pressure stripping (Gunn \& Gott 1972; Cortese et al. 2011), tidal interaction (Merritt 1983), galaxy harassment  (Moore et al. 1996) and viscous stripping (Rasmussen et al. 2012b). In addition to these studies which focus on the environment, there is also substantial work that is devoted in examining the properties of the HI sources themselves, like clustering and density, without focusing solely on individual galaxies (Chang et al. 2010; Masui et al. 2013; Switzer et al. 2013). Some of these studies focus on the particular environment of individual clusters which contain particular types of galaxies, such as late type galaxies or galaxies with specific morphology. Other studies focus on small galaxy samples without well defined selection criteria. So far, no coherent HI gas-galaxy-halo connections have been proposed.

In this paper, we carry out a systematic study for the HI gas-galaxy-halo connection using a large sample of local galaxy groups and the  HI blind survey ALFALFA data. We provide a model to describe the HI mass in galaxies, as a function of both galaxy properties and halo mass. In the near future, the next generation of HI surveys, such as the Five-hundred-meter Aperture Spherical radio Telescope (FAST, Nan et al. 2011); the on-going Australian SKA Pathfinder (ASKAP) survey; the Wide-field ASKAP L-Band Legacy All-Sky Blind Survey (WALLABY, Koribalski 2012) and the Westerbork Northern Sky HI Survey (WNSHS, Duffy et al. 2012), will be sensitive enough to detect faint HI emissions at high redshifts. These surveys will provide unprecedented opportunities to probe the HI gas-galaxy-halo connections, as well as the related galaxy formation processes.  The HI model described in this work can be used to construct mock HI catalogs and provide insight and predictions for the findings of these surveys.

The outline of this paper is as follows.  In section \S \ref{sec:data} we describe the HI source sample and galaxy group catalog used in our work.  In section \S \ref{sec:gal} we develop an HI-galaxy scaling relation, and used it as an initial try out model to predict the HI gas mass distribution. In section \S \ref{sec:halo} we investigate how halo mass can affect the total HI content within groups (halos), and construct the final galaxy/halo combined HI mass model. We support that the later can successfully predict various HI distribution properties. Finally, our results are summarized in section \S \ref{sec:summary}.

\begin{figure}
\vspace{0.0cm}
\center
\includegraphics[height=7.0cm,width=8cm,angle=0]{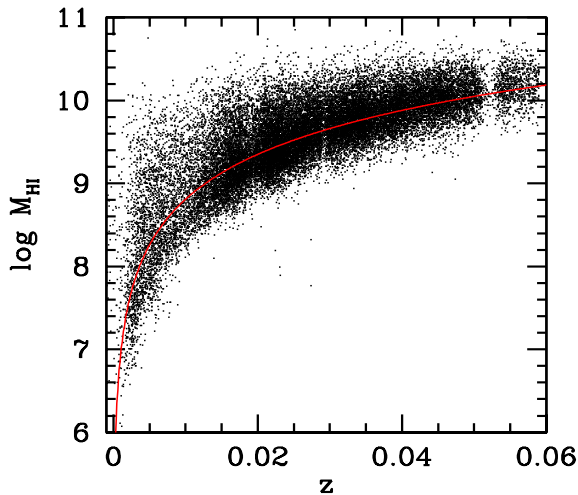}
\caption{The relation between the HI mass ${\rm \log M_{HI}}$ and the redshift $z$ for each ALFALFA HI source considered in this work. The red curve represents a fit to the maximum HI number counts in each HI mass bin as a function of redshift. The line roughly corresponds to the HI mass detection limit above which HI detection is complete.}
\label{fig:mhicomp}
\end{figure}

\begin{figure*}
\vspace{0.0cm}
\center
\includegraphics[height=5.0cm,width=10cm,angle=0]{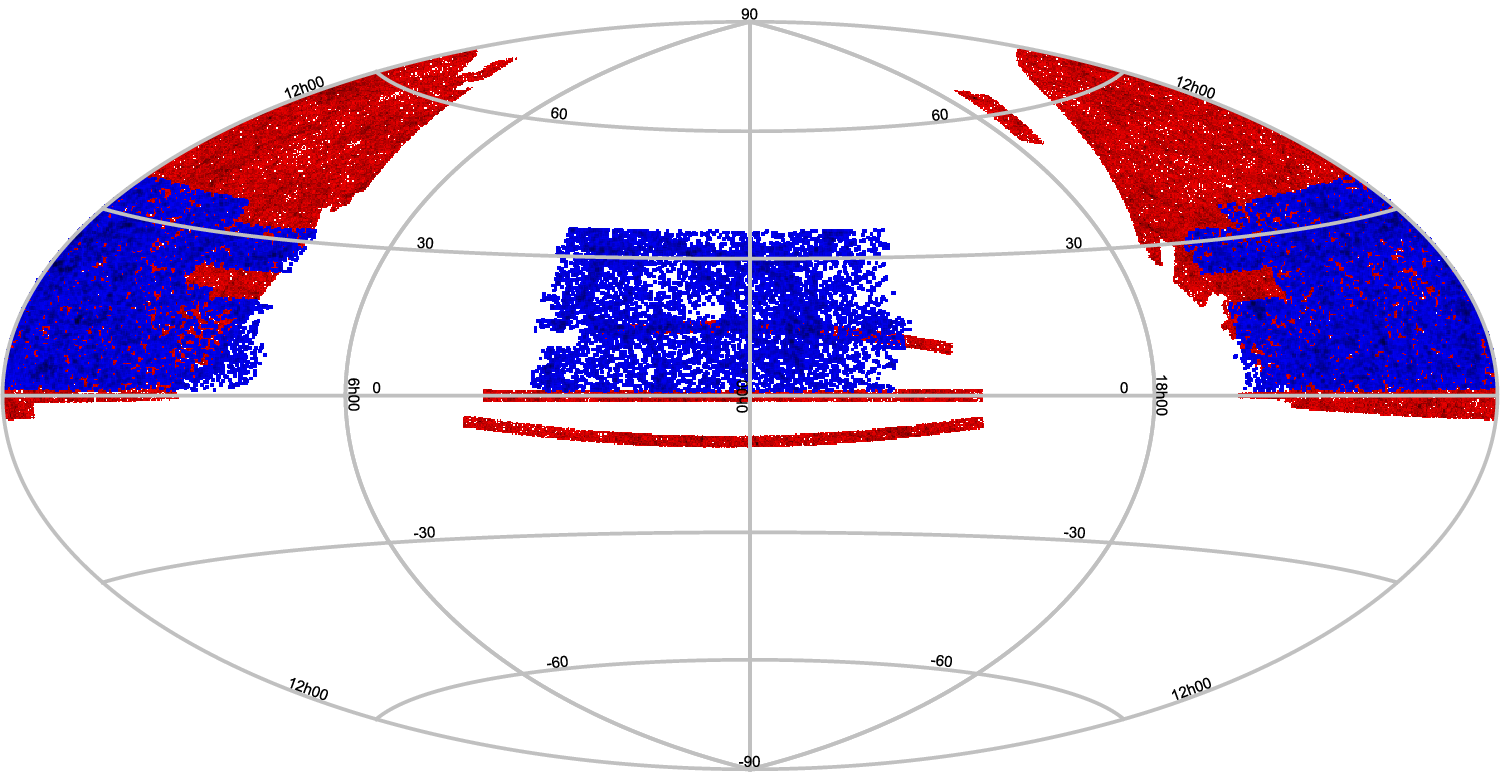}
\includegraphics[height=5.0cm,width=10cm,angle=0]{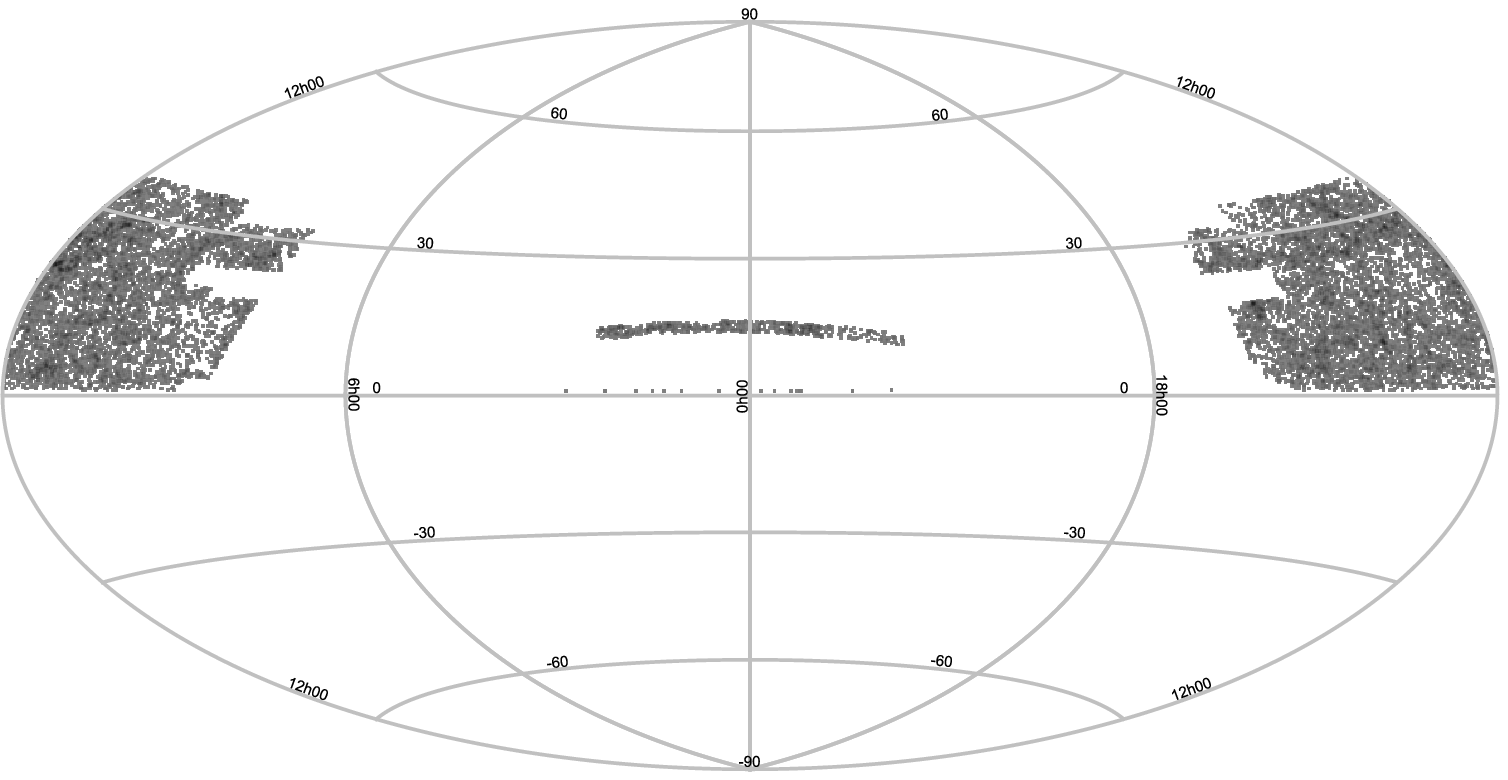}
\caption{Upper panel: the parent target distribution. The red dots represent the galaxies in the Y07 galaxy group catalog, while the blue dots show all the HI sources in ALFALFA catalog. Bottom panel: the HI-galaxy/groups pairs cross matched between the Y07 and the ALFALFA catalogs. }
\label{fig:Target}
\end{figure*}

\section[]{DATA}
\label{sec:data}

In this section, we describe the data used to constrain our HI gas modelling.

\subsection{The ALFALFA HI Sources}

The Arecibo Legacy Fast ALFA (ALFALFA, Giovanelli et al. 2005, Haynes et al. 2011) survey is a blind extra-galactic HI survey. It covers
approximately 7,000 ${\rm deg^2}$ on the north sky, including two separate regions: The first from $\sim 7.5 \, {\rm hr}$ RA to $\sim 16.5 \, {\rm hr}$ RA in the Arecibo Spring sky. The second from $\sim 22 \, {\rm hr}$ to $\sim 3 \, {\rm hr}$ RA in the Arecibo Fall sky. The blind observation of the 21-cm emission line is performed with the 305-m single-dish radio telescope at the Arecibo Observatory with an angular resolution of $3.5'$. In this study, we make use of the final data release (Haynes et al. 2018, hereafter $\alpha.100$), which contains 31,502 HI sources, up to redshift $z \sim 0.06$. Among the HI sources with available spectrum information, 31,158 sources ($99\%$) have optical counterparts, which are cross identified with the photometric and spectroscopic catalog associated with the SDSS. Within all these HI sources, there are 25,434 HI detections with secure extragalactic sources (labelled as "Code 1" in the ALFALFA catalog), and 6,068 sources categorized as "priors" (labelled as "Code 2" in the ALFALFA catalog). The later have low signal-to-noise ratio (${\rm S/N} \leq 6.5$) and usually are not considered reliable by the criteria set for "Code 1". The sources have been matched with their optical counterparts which are identified interactively by cross-matching external imaging databases (Haynes et al. 2011). In the ALFALFA catalog, each HI detection is characterized by its angular position in the sky, radial velocity, velocity width ${\rm W_{50}}$, and integrated HI line flux density ${\rm S_{21}}$. The HI mass (${\rm M_{HI}}$) is calculated via equation:
\begin{equation}\label{eq:mhi}
{\rm \frac{M_{HI}}{M_\odot}
=
2.356 \times 10^5 D^2_{Mpc}S_{21}\,,}
\end{equation}
following (Haynes et al. 2018).
${\rm D_{Mpc}}$ is the distance to the sources in ${\rm Mpc}$ and ${\rm S_{21}}$ is the integrated flux in ${\rm Jy \, km \, s^{-1}}$. We note that no correction for HI self-absorption has been applied.

Due to the survey's flux limit on the HI line flux ${\rm S_{21}}$, only a small portion of massive HI sources can be observed at a fixed  redshift, which in turn corresponds to a HI mass detection limit.  Fig. \ref{fig:mhicomp} shows the relation between the HI mass (${\rm \log M_{HI}}$) and redshift ${\rm z}$ for each ALFALFA HI source considered in this work. According to the HI mass function given by ALFALFA (Jones et al. 2018) the number density of HI mass bins increases with decreasing ${\rm \log M_{HI}}$, thus at a given redshift there are always more low mass HI sources than massive ones. The red curve in Fig. \ref{fig:mhicomp} represents a fit to the maximum HI number counts in each HI mass bin as a function of redshift. This line roughly corresponds to the HI mass detection limit above which HI detection is complete. There are totally  17,928 sources above this detection limit. Note that, the paucity of points at $\sim 0.053$ arises because many nights of ALFALFA observations have been contaminated by strong RFI generated by FAA radar at the San Juan airport (Haynes et al. 2011).

\subsection{Galaxy group catalog}

In this study, we make use of the SDSS galaxy group catalogs of Yang et al. (2007, hereafter Y07), constructed by employing the adaptive halo-based group finder of Yang et al. (2005). The parent galaxy catalog is the New York University Value-Added Galaxy catalog (NYUVAGC, Blanton et al. 2005), here updated to Data Release 7 (DR7, Abazajian et al. 2009), which contains an independent set of significantly improved reductions. The Main Galaxy group sample is constructed for galaxies in the DR7 complete to r-band apparent magnitude $r \sim 17.77$. The magnitudes and colors of all galaxies are based on the standard SDSS Petrosian technique. In this catalog, the $k+e$ corrected luminosities in the SDSS $ugriz$ bands and the stellar masses, estimated from the SDSS phorometry, are provided for each galaxy.

For each group included in the Y07 catalogue, the halo mass ${\rm M_h}$ is estimated by two methods. One is based on the ranking of the characteristic group luminosity, while the other is based on the ranking of the characteristic group stellar mass, which is defined as the total luminosity and stellar mass of all group members with $\rmag \leq -19.5$, respectively. Here the halo mass function obtained
by Tinker et al. (2008) adopts a WMAP7 cosmology with $A=200$, where ${\rm A}$ is the average mass density contrast in the spherical halo. The above two halo masses agree reasonably well with each other, while the difference between the two values decreases from $\sim 0.1$ dex at the low-mass end to $\sim 0.05$ dex at the massive end. For our work, we choose the ${\rm M_h}$ based on the ranking of group luminosity. For any  groups that the member galaxies are fainter than $\rmag = -19.5$, the halo masses are estimated according to the stellar to halo mass relation for central galaxies obtained in Yang et al. (2012). By default, we consider the brightest galaxy in a group as the central galaxy (BCG), while all the others are considered satellite galaxies. We also investigated the case in which the most massive galaxy in the group is regarded as the central galaxy (MCG). In most cases, these two definitions yield negligible consequences.

In summary, the group catalog used in our study contains 639,359 galaxies in the redshift range $0.01 \leq z \leq 0.20$, distributed in 472,416 groups within which about 23,700 have three member galaxies or more. The upper panel of Fig. \ref{fig:Target} displays the
ALFALFA HI detection sample and our SDSS  galaxy group catalog by blue and red dots separately. The low panel of Fig. \ref{fig:Target} demonstrates that the overlap between the ALFALFA and SDSS footprints is small in the southern Galactic cap (SGC) and quite big in the northern galactic cap (NGC).

\subsection[]{HI - galaxy group counterpart}

In the ALFALFA catalog, among the total 31,502 HI sources, 31,158 sources have been given optical counterparts by the ALFALFA team, which constitute about $99\%$ of the sample. SDSS images are used to identify interactively the most probable optical counterpart of each HI source. The resolution of the ALFALFA spectral grids is about $4'$, while the positions of the HI sources can be determined to an accuracy
typically better than $20''$. The identification of the optical counterparts is totally artificial and is based on information such as color, morphology, redshift and separation from the HI centroid. After processing each HI source, consistency checks are in order to evaluate redshift discrepancies or cases of large positional offset. More details of how to search and identify optical counterparts can be found in Haynes et al. (2011).

Among the SDSS galaxies identified as optical counterparts, 16,520 galaxies are included in the Y07 galaxy group catalog, corresponding to 14,270 groups. Within these 16,520 HI-galaxy pairs there are 12,892 HI-central galaxy pairs and 3,628 HI-satellite galaxy pairs. The  HI-galaxy pairs are selected as the target sample for our investigation. Note that we include both ALFALFA "Code 1" and "Code 2" sources, which signal-to-noise ratios are ${\rm S/N > 6.5}$ and ${\rm S/N \leq 6.5}$, respectively. All the HI-galaxy-group pairs considered  are illustrated in the bottom panel of Fig. \ref{fig:Target}. In this study, we use this sample to probe the {\it in situ} scaling relations between HI mass-galaxy properties and the {\it ex situ} influence of the larger scale halo environment.

We note that $\sim$30\% of HI detections appear at the overlap region of SDSS footprint and are identified to have optical counterparts to the ALFALFA survey, but their host galaxies are not included in the Y07 galaxy group catalog. In order to understand why the  Y07 galaxy group catalog missed the above HI matched galaxies, we matched the related HI detections with the SDSS DR13 galaxy catalog.
We find that $\sim$90\% of the objects are identified to have corresponding galaxies. However among these galaxies, only $\sim$45\%  have
 available spectra. Within the galaxies with available spectroscopic redshift information, $\sim$95\%  are fainter than the r-band magnitude limit of the Y07 galaxy group catalog. Thus, in short, the fact that most of the above HI optical counterparts are missed by the Y07 galaxy group catalog is due to two reasons:
\begin{itemize}
    \item They do not have redshift information.
    \item They are too faint to be included.
\end{itemize}

\section{HI mass estimator, try out Model 1}
\label{sec:gal}

In this section, we first focus on galaxies with available HI detections, i.e., the 16,520 HI-galaxy pairs. We start by investigating the impact of the HI-to-stellar mass ratio of these galaxies to their {\it in situ} galaxy properties (e.g. SFR and stellar mass) and present the HI mass estimator try out  Model 1. We pick the HI fraction ${\rm f_{HI}}$ as an indicator for the HI-to-stellar mass ratio
in a galaxy:
\begin{equation}\label{eq:fhig}
{\rm f_{HI}=\frac{M_{HI}}{M_s}\,,}
\end{equation}
where ${\rm M_{HI}}$ is the HI mass and ${\rm M_{s}}$ is the stellar mass.

\begin{figure*}
\vspace{0.0cm}
\center
\includegraphics[height=15.0cm,width=13cm,angle=0]{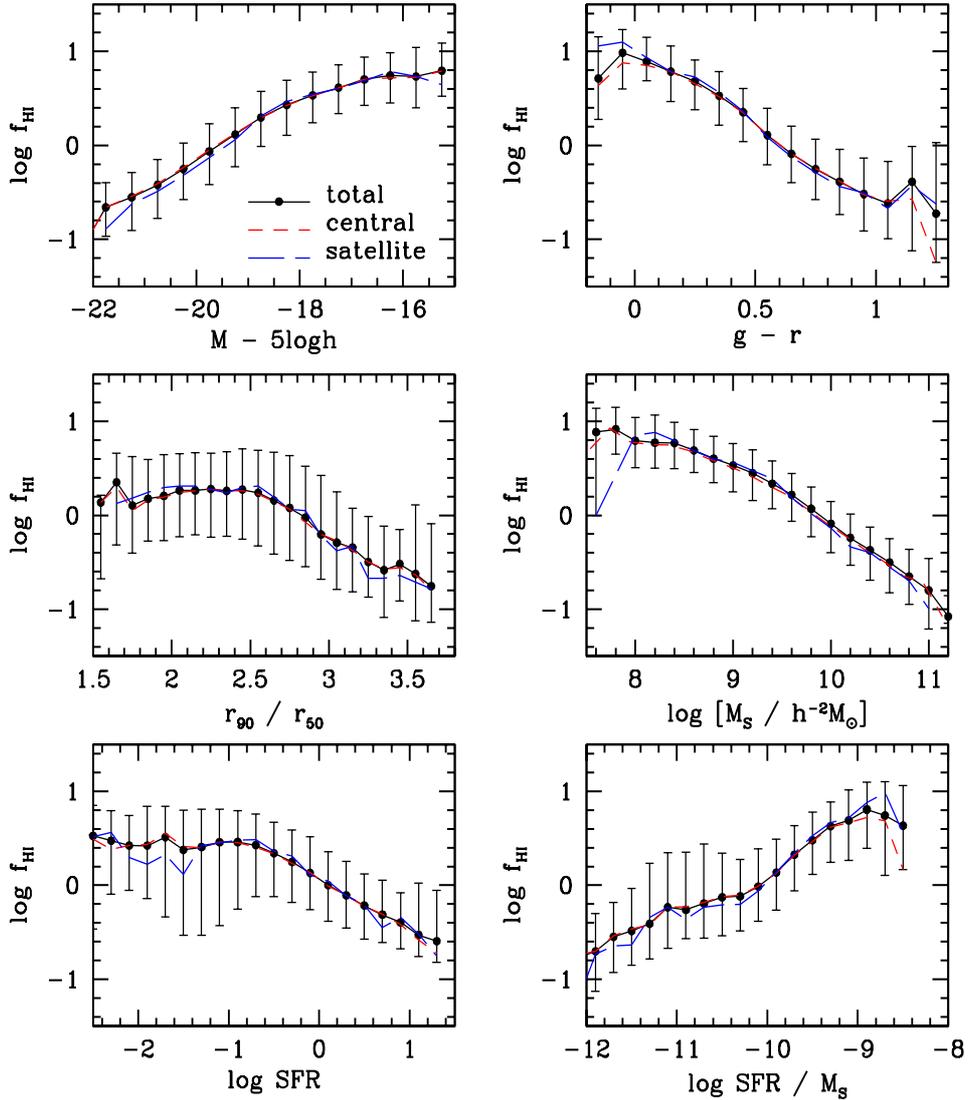}
\caption{The dependency of the HI mass fraction on  different galaxy properties for the matched HI-galaxy pairs. Top left panel: r-band absolute magnitude. Top right panel: color. Middle left panel: concentration. Middle  right panel: stellar mass. Bottom left panel: star formation rate. Bottom right panel: specific star formation rate. The black line represents the HI-galaxy pairs, while the red and blue dashed lines represent the central and the satellite galaxies, respectively. The error bars indicate the 68\% confidence level around the median.}
\label{fig:fracGal}
\end{figure*}

\subsection{HI fraction with respect to different {\it in situ} galaxy properties}
\label{sec:insitu}

Many studies have demonstrated that cold gas within galaxies is strongly related to key galaxy properties. Cold gas has been found to be correlated with galaxy stellar mass ${\rm M_s}$ (Cortese et al. 2011; Catinella et al. 2012; Huang et al. 2012),  optical color (Kannapan 2004b) and surface brightness (Zhang et al. 2009; Li et al. 2012). In this subsection we investigate how the HI faction in galaxies depends on galaxy properties in our sample.

In Fig. \ref{fig:fracGal} we present the HI fraction, ${\rm f_{HI}}$, of galaxies as a function of 6 key galaxy properties for the 16,520 HI-galaxy pairs cross matched between $\alpha.100$ and Y07. In the top left panel we present the dependence on the $r$-band absolute magnitude $\rmag$, at the top right panel on the ${\rm g - r}$ color, at the middle left panel on the concentration $r_{90}/r_{50}$, at the middle right panel on the stellar mass ${\rm M_s}$, at the bottom left panel on the star formation rate (SFR) and at the bottom right panel on the specific star formation rate (sSFR), defined as the ratio between star formation rate and galaxy stellar mass (i.e. ${\rm \log (SFR/M_s)}$). Overall, the HI fraction has quite strong dependence on the above galaxy properties, especially on stellar mass ${\rm M_s}$, color ${\rm g-r}$ and absolute magnitude ${\rm \rmag}$. Within the relations considered, it is not surprising that the HI fraction has a strong dependence on star formation rate (SFR) and specific star formation rate (sSFR) since the HI content provides the fuel to form stars. Next, the HI fraction also depends strongly on the galaxy concentration ${\rm r_{90}/r_{50}}$, where ${\rm r_{90}}$ and ${\rm r_{50}}$ are the radii containing $90\%$ and $50\%$ of the Petrosian flux in the ${\rm r}$-band. There is an obvious break at ${\rm r_{90}/r_{50} \sim 2.6}$, above which a strong anti-correlation between the HI fraction and concentration for galaxies emerges. The above can be explained by the fact that  the ${\rm r_{90}/r_{50} = 2.6}$ value is a threshold that divides early type galaxies from late types which may have been already quenched and have red colors (Strateva et al. 2001).

Among the relations considered some are clearly coupled with each other. For example, the absolute magnitude and stellar mass are strongly correlated with each other. Even concentration which is linked to the galaxy morphology, can be associated with color. The sSFR is by definition dependent on both SFR and stellar mass. According to the above couplings, we suggest that in order to account for the overall HI fraction in galaxies and construct a model, only two properties may be needed, e.g., color and stellar mass.  Indeed, the above  two parameters have already been proven to be the primary properties that are linked to the galaxy HI fraction by other authors (Bell et al. 2003; Zhang et al. 2009; Li et al. 2012).

In addition, since we have the central/satellite type information for the HI-galaxy pairs, it would be interesting to investigate if the two different populations demonstrate different galaxy property dependencies or not. To this end, we separate the HI-galaxy pairs as central and satellite subsamples. The results are represented by the red and the blue dashed lines of Fig. \ref{fig:fracGal}, respectively. The relations found for central and satellite galaxies do not show obvious differences with each other. Thus for our subsequent modeling for the HI gas component, we do not treat central and satellite galaxies separately.

\begin{figure}
\vspace{0.0cm}
\center
\includegraphics[height=11cm,width=8cm,angle=0]{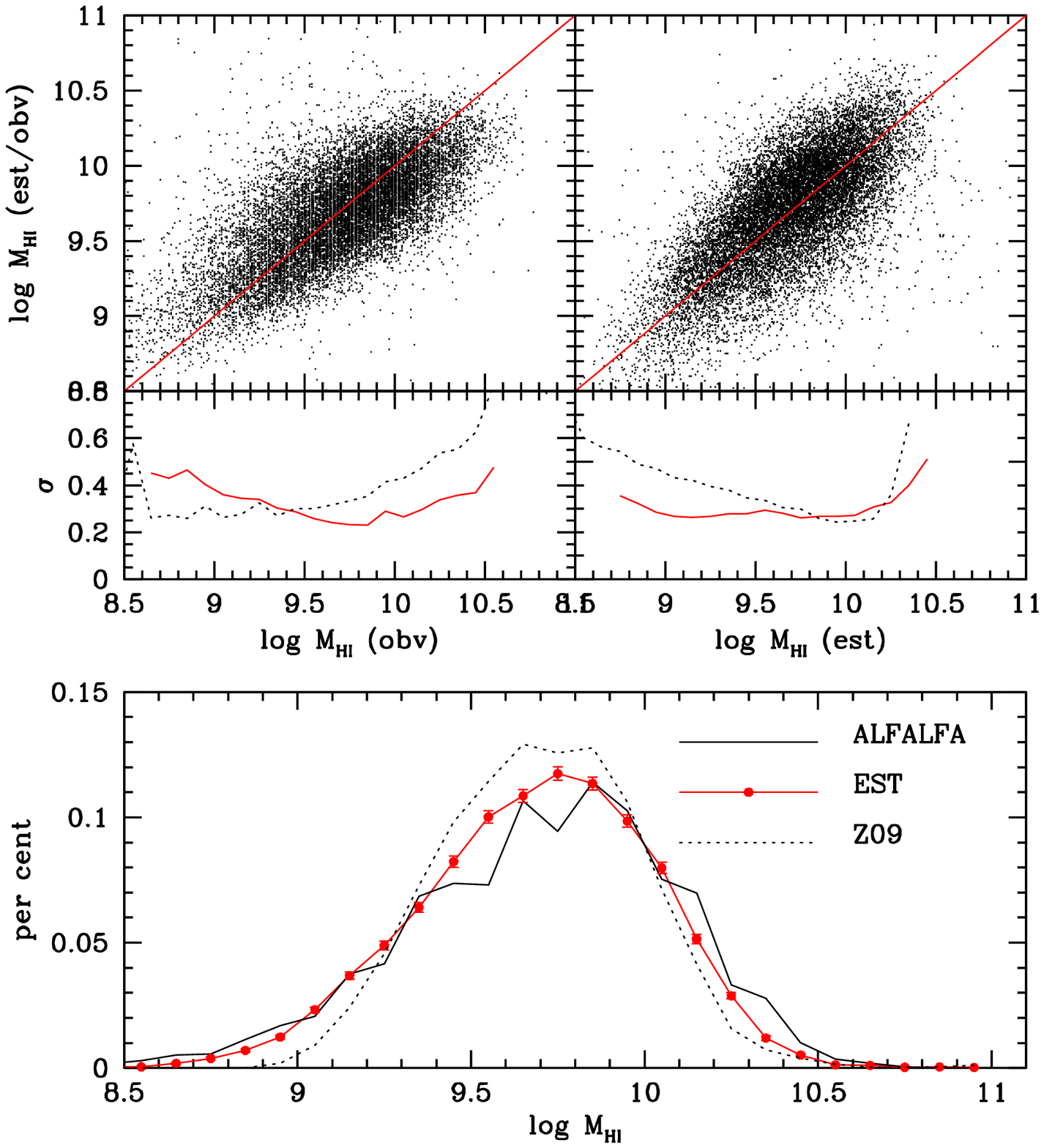}
\caption{Upper panel: The performance of our HI mass estimator (Model 1), i.e., the estimated HI mass v.s. the observed HI mass distributions. Middle panel: The standard variance given by our HI mass estimator (solid line) and  the one given by Z09  (dotted line). Lower panel: The HI mass distribution function given by $\alpha.100$ (black solid line), our model (red solid line) and Z09 (dotted line). }
\label{fig:estMhi}
\end{figure}

\subsection{The HI mass estimator Model 1}
\label{sec:estimator}

In this subsection we establish the HI mass estimator Model 1, following the relations given in the previous section for the HI-galaxy pairs. Other authors, have previously made attempts to calibrate the HI fraction to galaxy properties, trying to find an estimator (a combination of properties) of HI mass with a scatter as small as possible (Kannappan 2004a; Zhang et al. 2009; Li et al. 2012). Among the galaxy properties correlated with the HI mass fraction considered in our work, color (${\rm g-r}$) and stellar mass (${\rm M_s}$) are the most commonly used, since they strongly correlate with the HI content (subsection \ref{sec:insitu}), and also are easily accessible in  observational studies and mock observations (Katsianis et al. 2019,2020; Trvcka et al. 2020; Baes et al. 2020). Meanwhile, other properties considered, like the luminosity, star formation rate and concentration, basically are coupled with either stellar mass or color.

The strong correlations present in Fig. \ref{fig:fracGal}  imply that there potentially exist a plane with a well defined relation involving the following three variables: the HI mass fraction (${\rm f_{HI}}$), the stellar mass (${\rm M_s}$) and the color (${\rm g-r}$). In our work we assume a relation between the three properties of the following form:
\begin{equation}\label{eq:fHI}
{\rm \log f_{HI}=
a \log M_s + b (g-r) + c\,,}
\end{equation}
where the coefficients ${\rm a}$, ${\rm b}$, and ${\rm c}$ are free parameters. In order to infer the coefficients, we fit all the data points with the above equation using the Monte-Carlo Markov Chain (MCMC) method and explore the likelihood function in the multi-dimensional parameter space (Yan et al. 2003; van den Bosch et al. 2005) in our sample.
We give each data point a weighting coefficient ${\rm w}$ which is obtained by their signal-to-noise ratio (${\rm w = S/N/6.5})$. Thus,  we treat the HI source with higher ${\rm S/N}$ as more reliable than those with lower values. Then, we chose the parameter set which has the highest likelihood. The corresponding HI fraction can thus be described by the following equation,
\begin{equation}\label{eq:estgal}
{\rm \log f_{HI} = -0.224 \log M_s - 1.30 (g-r) + 2.96\,}.
\end{equation}
We adopt this relation as our try out model (Model 1 hereafter) for the HI mass estimation.

Before applying the above model to large samples, we first test its performance to our 16,520 HI - galaxy pairs, for which we already know the HI gas mass for each galaxy directly from the $\alpha.100$ catalog. Fig. \ref{fig:estMhi} shows the comparison between the HI mass estimated by our model (Eq. \ref{eq:estgal}), ${\rm M_{HI}(est)}$, and the observed HI mass provided by the $\alpha.100$ catalog, ${\rm M_{HI}(obv)}$. In order to quantify the scatter of the  ${\rm \log M_{HI}(est) = \log M_{HI}(obv)}$ relation which is represented by the red line in the top panel of Fig. \ref{fig:estMhi}, we measure the standard variance $\sigma$ between the estimated and the observed HI mass for each galaxy. The $\sigma$ is defined as:
\begin{equation}\label{eq: sigma}
{\rm \sigma = \sqrt{\frac{\sum_{i=1}^n(\log M_{est}-\log M_{obs})^2}{n-1}}}\,,
\end{equation}
where ${\rm n}$ is the number of galaxies in each HI mass bin with width ${\rm \Delta \log M_{obs}=0.1}$.  The result is shown in the middle panel of Fig. \ref{fig:estMhi} by the red curve. Overall, the scatter is close to $\sim 0.25$ dex for intermediate HI masses, increasing at the low and high mass end and demonstrating a U-shape form. At the bottom panel of Fig. \ref{fig:estMhi} we present the HI mass distributions. The observed distributions of HI mass given by $\alpha.100$ are demonstrated by a black curve while the estimated HI mass are represented by the red curve.

For comparison, we apply the HI mass estimator provided by Zhang et al. (2009, Z09 hereafter) to the same sample. The Z09 HI estimator is derived from 800 HI-galaxy pairs cross-matched using the SDSS DR4 and Hyper Leda HI data. The authors inferred the HI gas fraction for each galaxy using its color and surface stellar mass density $\mu_i$:
\begin{equation}\label{eq:MHI_Ms}
{\rm \log (M_{HI}/M_s)
=
0.215 \mu_i - 1.73 (g-r) - 4.08.}
\end{equation}
The standard variance ${\rm \sigma}$ and HI mass distribution given by the Z09 estimator are plotted in the middle and bottom panels and are represented by the black dotted curves. In general, our HI estimator has a similar performance with Z09. Our model performs better for predicting the HI mass galaxies at the high mass end, while Z09 tends to be more accurate at the low mass end.

\begin{figure}
\vspace{0.0cm}
\center
\includegraphics[height=14.0cm,width=8cm,angle=0]{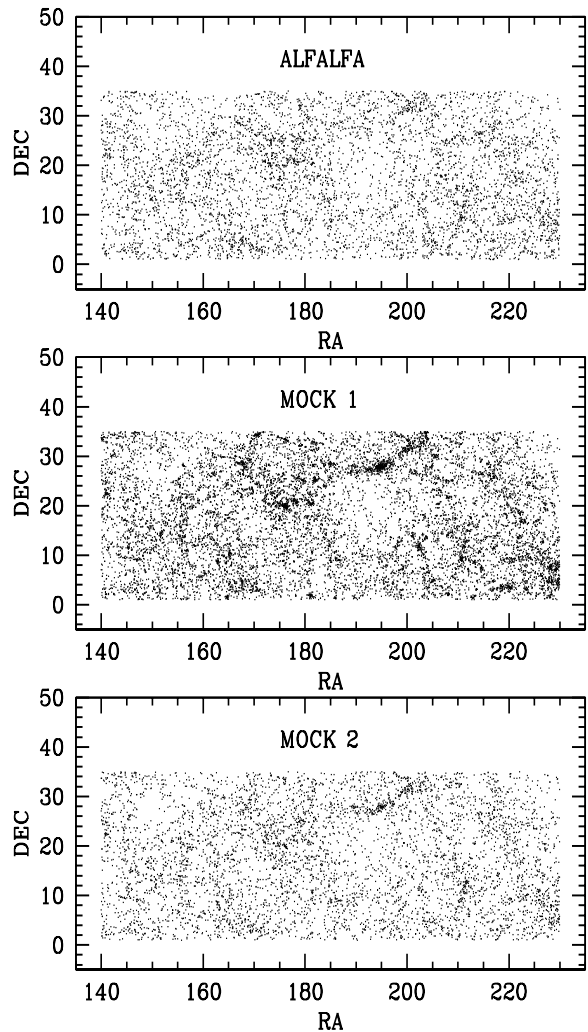}
\caption{The top panel shows the distribution of 5,668 detections given by the $\alpha.100$ catalog.  The middle panel shows the 11,084 detections predicted by our galaxy HI mass estimator (Mock 1). The bottom panel shows the 5,681 detections predicted by our halo based HI mass estimator (Mock 2).}
\label{fig:number}
\end{figure}

\subsection{Test Model 1 by constructing mock HI catalogs}
\label{sec:mockHI}

In order to test the performance of our try out Model 1, we directly apply it to the SDSS galaxy catalog used by Y07 and see if it  generates similar HI detections to the ALFALFA HI survey. The overlap of footprints between the Y07 and ${\rm \alpha.100}$ catalogs is mainly found for the NGC, where the redshifts of galaxies in the Y07 catalog ($z \sim 0.2$) extend to higher values. In order to compare with the $\alpha.100$ and eliminate effects related to redshift evolution, we only include galaxies with $z \leq 0.06$. Unlike the 16,520 HI - galaxy cross matched sample, most of these objects do not have HI detections. If our try out model is successful, i.e., the HI fraction depends on stellar mass and color (equation \ref{eq: sigma}), we should predict roughly the same HI detections as in the ALFALFA observations, given of course that all the observational selection effects are taken into account.

We first apply our HI mass estimator to all galaxies found in the Y07 which have $z \leq 0.06$. An HI mass ${\rm M_{HI}}$ is assigned for each galaxy using Eq. \ref{eq:estgal} and a scatter is associated, in accordance with the solid line shown in the middle panel of Fig. \ref{fig:estMhi}. These HI masses are all treated as potential HI sources. Due to the survey limit, not all of these HI sources would be detected. Thus, for the second step, we apply the same selection effects to the above sample as those of the ALFALFA and generate a mock HI catalog.

The ALFALFA is a blind and flux-limited HI survey. The survey depends both on the integrated HI line flux density ${\rm S_{21}}$ and the line profile width ${\rm W_{50}}$, as the detector is more sensitive to narrower line profiles than broader ones at a given ${\rm S_{21}}$. The flux density ${\rm S_{21}}$ is related to the HI mass (Haynes et al. 2011):
\begin{equation}\label{eq:S21}
{\rm \frac{M_{HI}}{M_{\bigodot}} = 2.356 \times 10^5 D^2_{Mpc} S_{21}\,,}
\end {equation}
while the line profile width is connected to the intrinsic rotational velocities
$v_{rot}$ by:
\begin{equation}\label{eq:vrot}
{\rm 2 v_{rot} = W_{50} / \sin (i)\,.}
\end {equation}
The galaxy inclination ${\rm i}$ is randomly selected from 0 to 90 degree and assigned to each galaxy in our catalog.
In addition, the rotational velocity correlates with the baryonic mass of a galaxy (McGaugh 2012),
\begin{equation}\label{eq:Mbvrot}
{\rm M_b = 47 V^4_{rot}\,.}
\end {equation}
Here, the baryonic mass ${\rm M_b}$ is the sum of all observed components, including stellar mass and gas (HI) mass:
\begin{equation}\label{eq:Mb}
{\rm M_b = M_s + M_{HI}\,.}
\end {equation}

Using Eqs. \ref{eq:S21},  \ref{eq:vrot}, \ref{eq:Mbvrot} and \ref{eq:Mb}, we obtain the velocity width of the HI line profile ${\rm W_{50}}$ in ${\rm km \, s^{-1}}$ and integrated HI line flux density of the source ${\rm S_{21}}$ in ${\rm Jy \, km \, s^{-1}}$ for each HI source predicted by our estimator. Whether a target can be observed or not by a galaxy survey depends on the completeness criterion, satisfying both the ${\rm W_{50}}$ and ${\rm S_{21}}$ limits. For the ALFALFA, a relationship between the ${\rm S_{21}}$ and the ${\rm W_{50}}$ of a source in terms of the signal-to-noise ratio ${\rm S/N}$ of the detection is given by Giovanelli et al. (2005):
\begin{equation}\label{eq:Slim1}
{\rm S_{21} = \left\{
\begin{array}{ll}
0.15S/N(W_{50}/200)^{1/2}, &  \mbox{$W_{50} < 200$}  \\
0.15S/N(W_{50}/200), &  \mbox{$W_{50} \geq 200$}
\end{array} \right.\,.}
\end {equation}
The above equation gives the expected theoretical survey completeness limit, which can be derived from the ALFALFA dataset. According to Saintonge (2007), the non-Gaussian noise of the automatic signal extractor for ALFALFA is generally above ${\rm S/N = 6.5}$. The authors assume that, for a flux-limited sample from a uniformly distributed population, the number counts will follow a power-law with an exponent of -3/2. Thus, the onset incompleteness can be determined when the data deviate from this form (Haynes et al. 2011). The resulting $90\%$ completeness limit for the ALFALFA sources can be expressed as:
\begin{equation}\label{eq:Slim2}
{\rm S_{21,lim} = \left\{
\begin{array}{ll}
0.5 \log W_{50} - 1.14, &  \mbox{$\log W_{50} < 2.5$}  \\
\log W_{50} - 2.39, &  \mbox{$\log W_{50} \geq 2.5$}
\end{array} \right.\,.}
\end {equation}

The distribution of the velocity width ${\rm W_{50}}$ versus the integrated flux density ${\rm S_{21}}$ plane based on the $\alpha.40$ catalog (Papastergis et al. 2011) shows that the detection limit of the survey is consistent with Eq. \ref{eq:Slim2}. Thus, we use the above to determine the detectable sources for our mock HI targets generated from Y07 galaxy catalog. We only include the potential HI targets with ${\rm S_{21} > S_{21,lim}}$. Meanwhile, other blind HI surveys, like HIPASS, have estimated the completeness of their catalogs as a function of the profile width ${\rm W_{50}}$. The distribution of profile widths ${\rm W_{50}}$ shows a cutoff at ${\rm 30 \, km \, s^{-1}}$ both for HIPASS and for ALFALFA. We also notice that measurements of the velocity width ${\rm W_{50}}$ extend up to ${\rm W_{50} \sim 20 \, km \, s^{-1}}$, which represents an additional survey limit for the catalog. Here, to comply with the above completeness limit, we also employ a limit of ${\rm W_{50,lim} = 20}$ for the  simulated HI targets from Y07. Last, we only select mock HI targets with ${\rm S_{21} > S_{21,lim}}$ and ${\rm W_{50} > W_{50,lim}}$ in order to comply with the $\alpha.100$ survey limit.

Once the mock HI catalog (hereafter Mock 1) under the ALFALFA survey criterion is constructed, we can proceed to compare it with the $\alpha.100$ observations within the same sky range and the same completeness limits. To ensure a fair comparison, we make the following constraints for both the observational and mock HI sources:

\begin{itemize}
\item We use a common volume by selecting an overlap sky coverage: ${\rm 140 < ra \leq 230}$ and ${\rm 1 < dec \leq 35}$, and an overlap redshift range: $0.01 < z \leq 0.06$.
\item In order to ensure that the surveys considered are complete in HI mass, we only consider objects in both samples with HI masses above the completeness limit represented by the red curve of Fig. \ref{fig:mhicomp}.
\item Galaxies within the Y07 catalog are only complete in terms of the ${\rm r}$-band apparent magnitude up to  ${\rm r = 17.77}$ due to limits of the SDSS survey. This means that many potential HI objects are neglected because their optical counterparts are too faint to be  detected. Thus, we remove the HI detections from the $\alpha.100$ sample which do not have optical counterparts in Y07.
\item  Since only bright galaxies can be observed and included in our analysis, the magnitude limit will induce a redshift dependent incompleteness in the number of galaxies and groups/halos. To make a fair comparison with ALFALFA that does not suffer from the halo mass incompleteness described above, we make use of the conservative halo-mass limit following Yang et al. 2009, i.e.:
\begin{equation}\label{eq:Mhlim}
{\rm \log M_{h,lim}
=
(z - 0.085)/0.069 + 12.0\,.}
\end{equation}
According to this criterion, at a given redshift ${\rm z}$, halos with masses larger than ${\rm M_{h,lim}}$ are considered to be complete within the survey. For both the mock and the observed HI detections, we choose to keep only the detections that are related to halos with masses larger than ${\rm M_{h,lim}}$.
\end{itemize}

With all these restrictions, we are set to compare the numbers and distributions of the HI detections from our HI estimator and $\alpha.100$ catalog. At the top and middle panels of Fig. \ref{fig:number} we present the distributions of these HI detections and compare the results of the observed $\alpha.100$ catalog and the Mock 1 HI catalog, respectively.  There are 5,668 detections for the $\alpha.100$ and 11,084 detections for our mock catalog. The mock catalog contains $\sim 95\%$ objects more than the observed one. An additional test using other HI estimators, like the one given by Z09, also shows that the HI detections are overestimated significantly. A reasonable explanation is that there is a large fraction of galaxies which do not follow the HI mass scaling relation derived from the observed HI-galaxy pairs and the try out model (and similar calibrations) does not perform adequately.

\section[]{HI mass estimator final model 2}
\label{sec:halo}

In section \ref{sec:mockHI} we demonstrated that our Mock 1 catalog significantly over-predicts the total HI population. The above implies that either 1) we have assigned galaxies with too high HI masses, or 2) that the large scale environment, e.g., the halo, is able to deplete  somehow the HI gas. To test the first assumption, we separate the HI-galaxy pairs into two subsamples, one that contains galaxies with redshift $z\ge 0.03$ and the other $z<0.03$. Since the latter subsample suffers less from the ALFALFA flux limit, the HI mass distributions as well as the scaling relation that is used to  derive the HI mass should shift to lower HI mass values. However, we do not see such a trend. Thus, we decide to explore the second assumption, i.e. the hypotheis that the HI gas in galaxies is significantly impacted by the halo environment. A question that arises is : environment is able to {\t reduce} HI sources in each halo ? or {\it totally remove all} the HI sources in certain halos ? In order to quantify such an environment effect, we set out to measure the total HI gas within each galaxy group (halo).

\begin{figure}
\vspace{0.0cm}
\center
\includegraphics[height=14cm,width=8cm,angle=0]{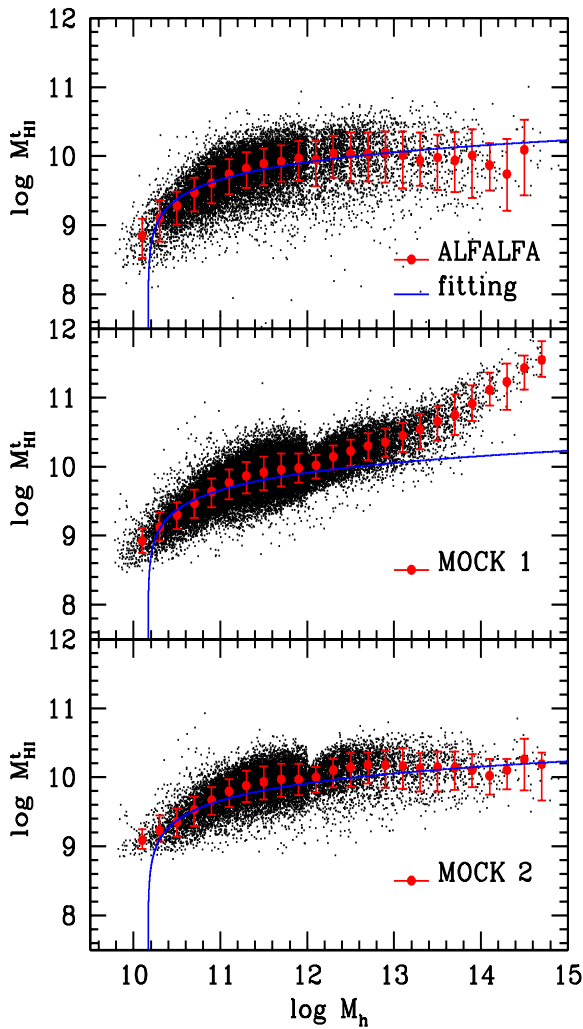}
\caption{The relation between the total HI mass and halo mass. Top panel: The observed ${\rm M_{HI}^t}$-${\rm M_{h}}$ relation. Middle panel: The ${\rm M_{HI}^t}$-${\rm M_{h}}$ from the Mock 1 catalog. Bottom panel: The ${\rm M_{HI}^t}$-${\rm M_{h}}$ from the Mock 2 catalog.   Black points represent the groups that have HI detections, while the red symbols with error bars represent the median and  68\% confidence level for all the data. The blue solid line represents a fit to the observed data from ALFALFA.}
\label{fig:MhitMh}
\end{figure}

\subsection[]{Total HI mass in halos}

In this subsection, we measure the total HI gas within galaxy groups. Instead of the ${\rm M_{HI}}$ of individual galaxies, here we focus on the ${\rm M_{HI}^t}$, defined as the total cold gas mass within each halo. This is equal to the sum of the HI masses of all HI detections located within each separate group (halo). Halo mass is a proxy of the halo environment and is also included in the Y07 group catalog. At the top panel of Fig. \ref{fig:MhitMh}  we present the relation between the ${\rm M_{HI}^t}$ and halo mass ${\rm M_h}$ for the cross matched objects between the Y07 and $\alpha.100$ catalogs, after taking into account the completeness limit cuts as outlined in section \ref{sec:mockHI}. The red points with error bars represent the median and $68\%$ confidence level at each halo mass bin, respectively.

It is expected that the total HI mass ${\rm M_{HI}^t}$ increases with halo mass, since more massive halos tend to contain more member galaxies and thus more cold gas. However, in Fig. \ref{fig:MhitMh} we demonstrate that the relation between ${\rm M_{HI}^t}$ and ${\rm M_h}$  is not linear. Interestingly, over a very large mass range $({\rm \log M_h \geq 11.0})$, the total HI mass is almost unchanged. We remind that our estimation of HI masses is based on a flux-limited HI survey. Including fainter HI sources, below the HI survey limit might slightly increase the total HI mass within different halos. However, since we focus on the total HI mass as a function of halo mass, not as a function of redshift, the incompleteness factor is not expected to change the slope of the ${\rm M_{HI}^t - M_h}$ relation significantly. We perform a fit to the ${\rm M_{HI}^t - M_h}$ distribution to obtain the following best fit equation:
\begin{equation}\label{eq:MHI}
{\rm \log M_{HI}^t
= 9.7 (\log M_h -10.2)^{0.03}\,.}
\end {equation}
The best fit is plotted at the top panel of Fig. \ref{fig:MhitMh} as a blue curve.

We investigate the total HI mass as a function of halo mass for our Mock 1 catalog at the middle panel of  Fig. \ref{fig:MhitMh}.
For comparison, we also include the best fit line obtained from the observational data from ALFALFA via the  blue solid line. With respect  the observations, the total HI mass in massive halos is significantly enhanced in the Mock 1 catalog. The comparison suggests that large halos should undergo somehow an HI deficit.

\begin{figure}
\vspace{0.0cm}
\center
\includegraphics[height=10.0cm,width=8cm,angle=0]{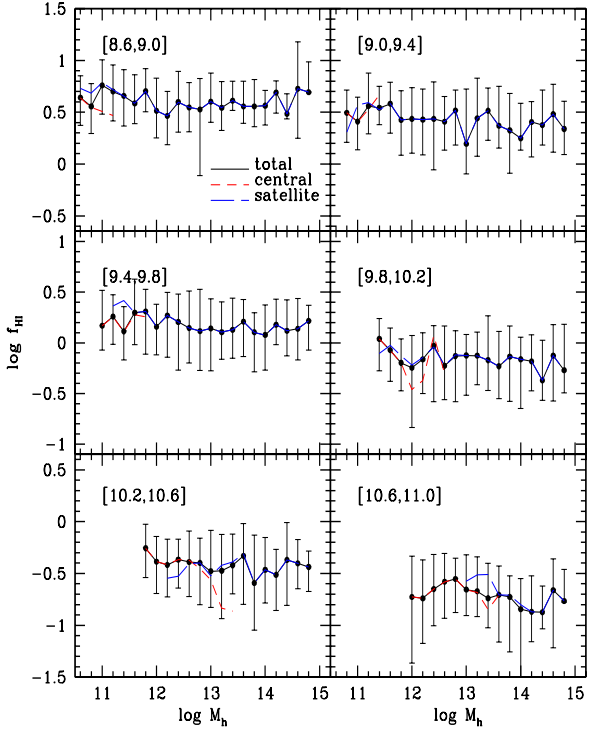}
\caption{The relation between galaxy HI mass fraction and halo mass ${\rm M_h}$ at different mass scales. Top left panel: ${\rm M_s} = 10^{8.6-9.0}$ ${\rm M_{\odot}}$, top right panel: ${\rm M_s} = 10^{9.0-9.4}$ ${\rm M_{\odot}}$, middle left panel: ${\rm M_s} = 10^{9.4-9.8}$ ${\rm M_{\odot}}$ , middle right panel: ${\rm M_s} = 10^{9.8-10.2}$ ${\rm M_{\odot}}$,  bottom left panel: ${\rm M_s} = 10^{10.2-10.6}$ ${\rm M_{\odot}}$, bottom right panel: ${\rm M_s} = 10^{10.6 - 11.0}$ ${\rm M_{\odot}}$. The black line represents all HI-galaxy pairs, while the red and blue dashed lines represent the central and satellite galaxies, respectively. The error bars indicate the 68\% confidence level around the median.}
\label{fig:fracMs}
\end{figure}

\begin{figure}
\vspace{0.0cm}
\center
\includegraphics[height=10.0cm,width=8cm,angle=0]{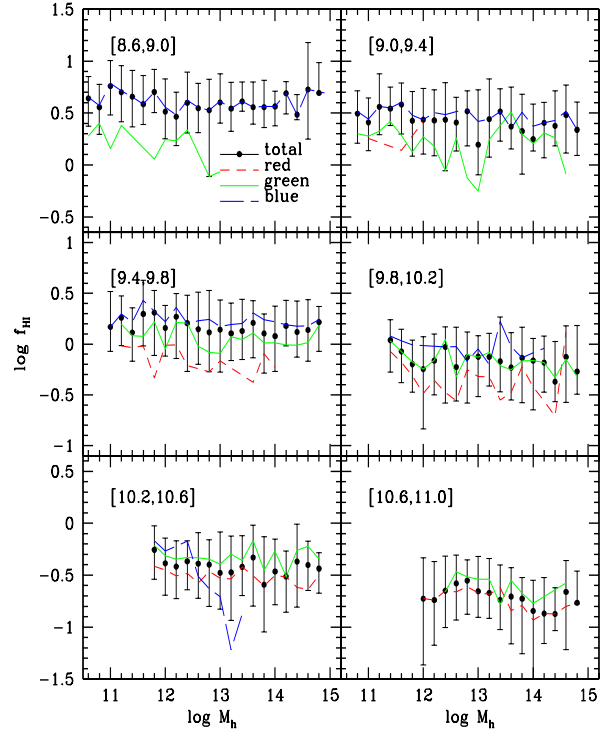}
\caption{Similar to Fig.\ref{fig:fracMs} but galaxies are separated into different color sub-samples.}
\label{fig:fracColor}
\end{figure}

\subsection{Instant or continuous depletion?}
\label{sec:exsitu}

After demonstrating the presence of an HI deficient scenario in halos, we proceed to address the following question: The HI depletion is an instant or a continuous process? To answer this question, we probe the HI fraction of the HI-galaxy pairs as a function of halo mass. Since many observations suggest that the interaction between galaxies and intracluster medium is important for removing/heating gas from galaxies. If the HI depletion is a continuous process, we shall find a halo mass dependence since galaxies in more massive halos suffer from stronger depletion effect. In order to disentangle the dependence of the HI fraction on other galaxy  properties, we separate the sample into different sub-samples according to their stellar mass and color. In Fig. \ref{fig:fracMs} we present the HI fraction ${\rm f_{HI}}$ of galaxies for sub-samples in  six stellar mass bins (top left panel: ${\rm M_s} = 10^{8.6-9.0}$ ${\rm M_{\odot}}$, top right panel: ${\rm M_s} = 10^{9.0-9.4}$ ${\rm M_{\odot}}$, middle left panel: ${\rm M_s} = 10^{9.4-9.8}$ ${\rm M_{\odot}}$ , middle right panel: ${\rm M_s} = 10^{9.8-10.2}$ ${\rm M_{\odot}}$,  bottom left panel: ${\rm M_s} = 10^{10.2-10.6}$ ${\rm M_{\odot}}$, bottom right panel: ${\rm M_s} = 10^{10.6 - 11.0}$ ${\rm M_{\odot}}$). We find that there is no significant halo mass dependence.
In addition, we separate galaxies into centrals (red dashed line) and satellites (blue dashed line). We do not find obvious differences between centrals and satellites, while there is no dependence of the ${\rm f_{HI}}$ on halo mass. Furthermore we also separate the galaxies into different color subsamples and present the results in Fig. \ref{fig:fracColor} for red (${\rm g-r \geq 0.8}$), green (${\rm 0.5 \leq g-r < 0.8}$) and blue (${\rm g-r < 0.5}$) galaxies, respectively.  We can see that galaxies of different colors, despite the fact that they  overall have different average values for ${\rm f_{HI}}$, have fractions with no significant halo mass dependence.

In Fig. \ref{fig:fracMs} and Fig. \ref{fig:fracColor} we demonstrate that the HI fraction of the  observed HI-galaxy pairs does not show  significant halo mass dependence. Thus, we assume that the HI gas depletion in dark matter halos is an instant and not a continuous process. We note that once a depletion effect takes place, it instantly depletes the HI gas associated with the galaxy and thus we do not observe this HI-galaxy pair anymore.

\begin{figure}
\vspace{0.0cm}
\center
\includegraphics[height=6.5cm,width=8cm,angle=0]{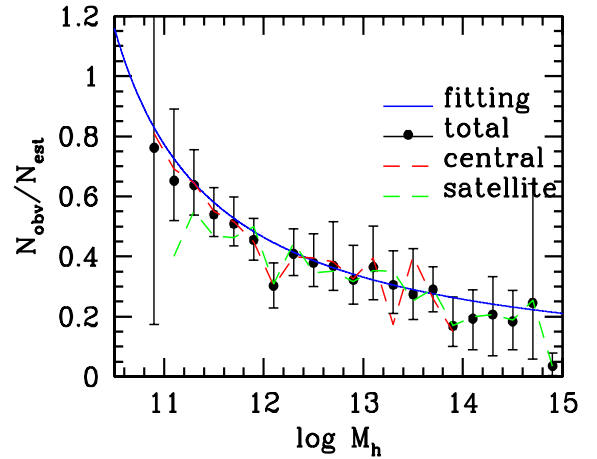}
\caption{ The relation between halo mass and the HI survive efficiency defined as the ratio between observed HI detections ${\rm N_{obv}}$ and mocked HI detections by galaxy HI estimator ${\rm N_{est}}$ within the halo. The error bars are given by 500 bootstrap re-samplings.}
\label{fig:efc}
\end{figure}

\subsection[]{The HI mass estimator Model 2}

To properly model the impact of a halo to the HI gas of its member galaxies, we introduce a concept for the survival of HI mass labelled as the  `efficiency' factor, which describes the probability of a halo of {\it not} removing any of its HI gas component. The above is defined as the ratio of the HI detections which have been observed (${\rm N_{obv}}$) via the detections which would be observed (${\rm N_{est}}$) according to Model 1, i.e. ${\rm N_{obv}/N_{est}}$. More specifically ${\rm N_{obv}}$ represents the number of HI detections at the ${\rm \alpha.100}$ catalog, while ${\rm N_{est}}$ represents the detections for the Mock 1 model (Eq. \ref{eq:estgal}). Note that here we apply the same survey selection criteria  for both the observed and Mock1 catalogs, including the halo mass completeness limit, as described in section \ref{sec:estimator}. Ideally, if the halo environment would not have any impact on the HI component of galaxies we would have  ${\rm N_{obv}/N_{est} = 1}$. But as shown in section \ref{sec:mockHI}, it is expected that the ${\rm N_{est}}$ is much larger than the ${\rm N_{obv}}$ and thus ${\rm N_{obv}/N_{est}}$ is expected to be lower than 1.

Fig. \ref{fig:efc} shows the `efficiency' for halos at different mass scales. Black symbols with error bars represent the overall population ratio ${\rm N_{obv}/N_{est}}$ as a function of halo mass. Here the error bars are obtained from 500 bootstrap re-samplings. We also investigate seperately the central and satellite populations by following the red and green dashed curves, respectively. Overall, the `efficiency' is decreasing with increasing halo mass. The above suggests that the ability of halo environment on suppressing its HI gas component is stronger in more massive halos. The picture persists for central and satellite galaxies, except at the low mass end. We note that in our analysis, due to the magnitude limits of the survey, the total number of satellite galaxies is small and thus the uncertainty at the last mass bin is quite large.

In order to better quantify the HI survival efficiency, we perform a fit to the data shown in Fig. \ref{fig:efc} and obtain the following relation:
\begin{equation}\label{eq:efc}
{\rm f_e(M_h)=N_{obv}/N_{est}=
1.16 (\log M_h -9.5)^{-1.0}\,.}
\end {equation}
The best fit line is shown in Fig. \ref{sec:mockHI} (dashed blue line). We note that Zhang et al. (2020)  found that the ${\rm H_{\alpha}}$ emission in groups demonstrates an `efficiency' factor with a similar behavior.

Using the efficiency factor described above, it is quite straightforward to construct a final HI mass estimator model (Model 2), which has two components: one related to galaxies and one related to halos. Since we take into account both the properties of the galaxy and its parent halo to construct our HI model, we refer to the above as the `halo-based' HI estimator (Model 2). It contains the following two steps:

\begin{itemize}
\item First, similar with model 1 described in section \ref{sec:estimator}, we treat all galaxies in Y07 as potential HI source candidates. We assign each object an HI mass according to our galaxy HI mass estimator (Eq. \ref{eq:estgal}), with the scatter taken account in the model.

\item Second, according to the host halo mass of that galaxy, we get an HI survival `efficiency' factor using Eq. \ref{eq:efc}. We draw a random value ${\rm f_i}$ equally distributed within [0,1]. If ${\rm f_i\le f_e}$, we keep the resulting HI mass. Otherwise, the HI mass is set to be zero.
\end{itemize}

We assume that the amount of the HI content within a galaxy is primary depend on the galaxy properties. However, the halo environment can decide whether a galaxy has cold gas or not. Many previous studies have shown that galaxies, regardless if they have an HI detection or not, do not have a significant statistically difference in their properties. This feature is also confirmed in section \ref{sec:exsitu}. Thus, we support that our assumption is reasonable.

\begin{figure}
\vspace{0.0cm}
\center
\includegraphics[height=16.0cm,width=8cm,angle=0]{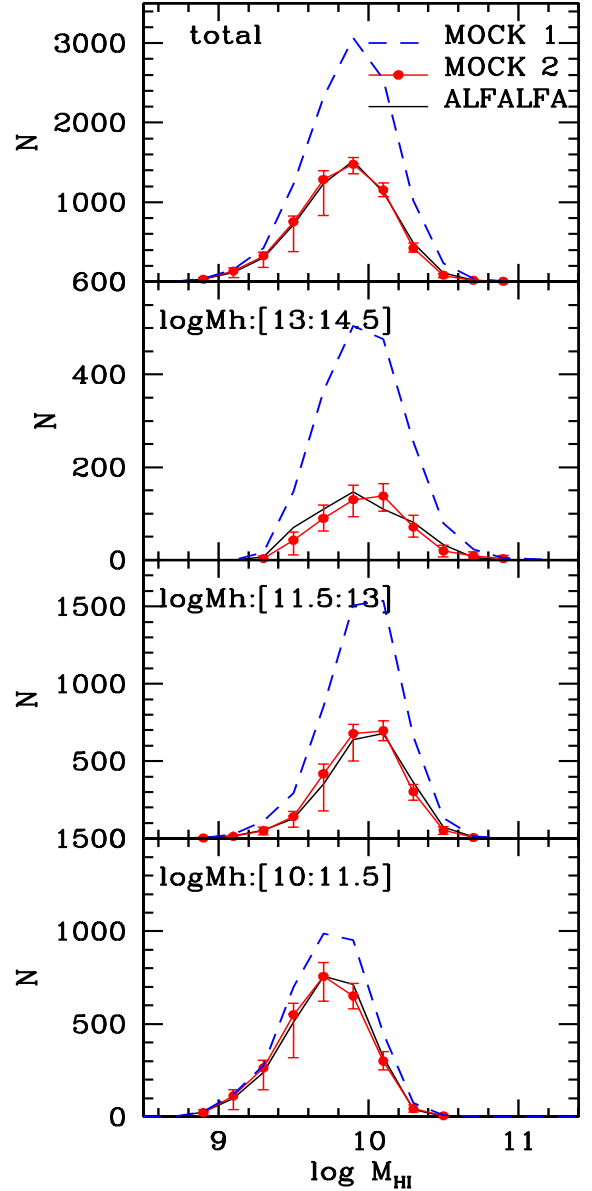}
\caption{A comparison between the estimated halo HI mass distribution and the observational HI mass distribution for different halo mass
environment. The dots with errorbars, the solid black and the dashed blue lines represent the Mock 2, ALFALFA and Mock 1 catalogs, respectively. The error bars are given by 500 bootstrap re-sampling. }
\label{fig:distrimhi_bin}
\end{figure}

\begin{figure*}
\vspace{0.0cm}
\center
\includegraphics[height=6.0cm,width=13cm,angle=0]{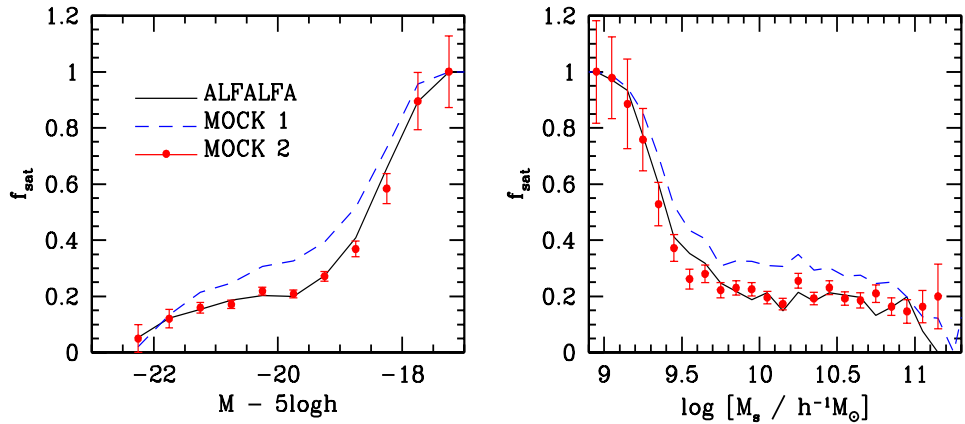}
\caption{The fraction of satellite galaxies ${\rm f_{sat}}$ as a function of absolute magnitude (left panel) and stellar mass (right panel). The dots with errorbars, the solid black and tue dashed blue lines represent Mock 2, ALFALFA and Mock 1 catalogs, respectively. }
\label{fig:fsat_mock}
\end{figure*}

\subsection[]{The performance of our final Model 2}

To assess our `halo-based' HI mass estimator (final Model 2), we apply it to the SDSS observation and get an updated mock HI catalog (hereafter Mock 2). Similar to the Mock 1 HI catalog, we also apply various survey selection limits to it. We compare the Mock 2 HI detections with the real HI detections provided by the ${\rm \alpha.100}$ catalog under the same completeness cuts.

The bottom panel of Fig. \ref{fig:number} displays the distribution of the Mock 2 HI detections generated by our halo based HI model. For  this sky coverage, there is a total of 5,681 detections. In the top panel, we compare the predictions from the Mock 2 model to the observations obtained from the ${\rm \alpha.100}$ catalog for the same sky range. We find that there is only a difference of  $0.2\%$ in the number of HI sources. At the middle panel of the same figure we demonstrate that the visual distribution pattern obtained for the Mock 2 catalog is much closer to the pattern obtained in observations.

As a consistency check, at the lower panel of Fig. \ref{fig:MhitMh} we present the total HI mass - halo mass relation generated for our Mock 2 HI catalog and demonstrate that it is consistent with the one obtained from the ${\rm \alpha.100}$ data. Our halo-based HI estimator predicts a similar HI mass - halo mass relation.

As an independent assessment for the goodness of our Model 2, we compare the HI mass distributions of observation and mock samples. The above comparison is shown at the upper panel of Fig \ref{fig:distrimhi_bin}. The black curve represents the ${\rm \alpha.100}$ catalog, while the blue and red curves are given by the Mock 1 and the Mock 2 catalogs, respectively. Here all HI detections are restricted by the same sky coverage and under the same completeness criteria. Overall, the distributions given by the Mock 2 model and observation are in good agreement with each other, while the Mock 1 model significantly over-predicts the results.

An other indicator of the success of our model comes from the comparison between the observed and the simulated number distributions of HI sources at different halo mass bins (three bottom panels of Fig. \ref{fig:distrimhi_bin}).  We can see that over different halo mass scales, the Mock 2 catalog shows a very good agreement with the observational data. The Mock 1 catalog demonstrates a large excess in numbers with respect observations, while the limitation of the model becomes more severe for more massive halos.

As a final test, we compare the satellite fractions of galaxies predicted by our Mock 2 catalog to those obtained from the Mock 1 catalog and the observed HI-galaxy pairs (Fig. \ref{fig:fsat_mock}). We note that in order to make a fair comparison between the above, we apply the selection criteria described in section \ref{sec:mockHI}. Because we applied these criteria, the satellite fractions shown in this plot are quite different from those obtained by Yang et al. (2008), especially at the low mass end, since faint central galaxies are not included in our analysis. Model 2 performs much better than Model 1 in replicating the observed ${\rm \alpha.100}$ satellite fraction as a function of stellar mass and luminosity.

We plan to demonstrate in a companion paper that our model is able to predict the correct HI mass function and clustering properties (Wang et al. 2020, in preparation). All these tests demonstrate that our halo-based HI estimator captures important features for the HI-galaxy-halo connection. The model can also be applied to predict HI catalogs for deeper HI surveys like FAST.

\section[]{SUMMARY}
\label{sec:summary}

In this study, we have constructed a `halo-based' HI mass estimator, employing the relation of HI sources with their {\it in situ} galaxy
and {\it ex situ} host halo environments. The analysis makes use of 16,520 HI-galaxy pairs extracted from the ${\rm \alpha.100}$ catalog which correspond to 14,270 HI cross matched groups from the SDSS DR7 galaxy group catalog (Y07). Our main results are summarized bellow:

\begin{itemize}

\item There is a strong dependence of a galaxy HI fraction with properties like absolute magnitude, color, stellar mass, star formation rate and concentration. By employing such relations we develop a galaxy HI mass estimator (try out Model 1) for the {\it detected} HI-galaxy pairs. Our estimator involves the ${\rm g-r}$ colors and stellar masses of galaxies, while it can predict the HI mass of a galaxy with a standard variance of $\sigma \sim 0.25$.

\item Applying our try out Model 1 to the SDSS DR7 galaxy catalog, and taking into account various selection completeness limits, we demonstrated that this galaxy-related HI mass estimator significantly overestimates HI detections with respect the observational results.

\item  By comparing the total HI mass in groups (halos) to the observed and mock HI catalogs, we found an obvious suppression of the total HI content within groups with halo masses of $\log M_h \geq 12.0$.

\item The total HI mass in a group obtained from the ALFALFA observations keeps almost unchanged with increasing halo mass for $\log M_h \geq 11.0$ objects.

\item  Based on the above findings, we proposed a `halo-based' HI mass estimator (final Model 2), which involves the ability of the HI gas (mass) to survive in halos of different masses, by taking into account an additional HI survive `efficiency' factor.

\item Applying our `halo-based' HI mass estimator (Model 2) to the SDSS DR7, we generate a mock HI catalog that is in good agreement with  observations in terms of the total number, HI mass distribution, total HI in halos and satellite fractions.

\end{itemize}

We support that our model is ready to be applied to galaxy surveys (like FAST and SKA) to predict the HI gas distribution of the local Universe. Meanwhile, our HI-halo efficiency model may contain important information about the interaction of galaxies and underlying feedback in dark matter halos. We will investigate the above using Hydro-dynamical simulations in future work.

\section*{Acknowledgements}

This work is supported by the national science foundation of China
(Nos. 11833005,  11890692, 11621303), 111 project No. B20019 and
Shanghai Natural Science Foundation, grant No. 15ZR1446700, 19ZR1466800.
We also thank the support of the Key Laboratory for Particle
Physics, Astrophysics and Cosmology, Ministry of Education and Tsung-Dao Lee Institute.

\bibliographystyle{mnras}	

\begin{thebibliography}{}

\bibitem[Abazajian et al. (2009)]{Abazajiann2009} Abazajian, K. N., Adelman-McCarthy, J. K., Agueros, M. A., et al.\ 2009, \apj S, 182, 543

\bibitem[Baes et al.(2020)]{2020MNRAS.494.2912B} Baes, M., Tr{\v{c}}ka, A., Camps, P., et al.\ 2020, \mnras, 494, 2912
    
\bibitem[Bell et al.(2003)]{2003ApJS..149..289B} Bell, E.~F., McIntosh, D.~H., Katz, N., et al.\ 2003, \apjs, 149, 289
    
\bibitem[Berlind \& Weinberg(2002)]{2002ApJ...575..587B} Berlind, A.~A. \& Weinberg, D.~H.\ 2002, \apj, 575, 587

\bibitem[Berlind et al.(2006)]{2006ApJS..167....1B} Berlind, A.~A., Frieman, J., Weinberg, D.~H., et al.\ 2006, \apjs, 167, 1
    
\bibitem[Blanton et al.(2005)]{2005ApJ...629..143B} Blanton, M.~R., Eisenstein, D., Hogg, D.~W., et al.\ 2005, \apj, 629, 143
    
\bibitem[Boselli et al.(2001)]{2001AJ....121..753B} Boselli, A., Gavazzi, G., Donas, J., et al.\ 2001, \aj, 121, 753

\bibitem[Brown et al.(2016)]{2016MNRAS.463.2839B} Brown, G.~M., Johnston, K.~G., Hoare, M.~G., et al.\ 2016, \mnras, 463, 2839

\bibitem[Catinella et al.(2010)]{2010MNRAS.403..683C} Catinella, B., Schiminovich, D., Kauffmann, G., et al.\ 2010, \mnras, 403, 683

\bibitem[Catinella et al.(2012)]{2012A&A...544A..65C} Catinella, B., Schiminovich, D., Kauffmann, G., et al.\ 2012, \aap, 544, A65

\bibitem[Chang et al.(2010)]{2010Natur.466..463C} Chang, T.-C., Pen, U.-L., Bandura, K., et al.\ 2010, \nat, 466, 463
    
\bibitem[Cortese et al.(2008)]{2008MNRAS.383.1519C} Cortese, L., Minchin, R.~F., Auld, R.~R., et al.\ 2008, \mnras, 383, 1519
    
\bibitem[Cortese et al.(2011)]{2011MNRAS.415.1797C} Cortese, L., Catinella, B., Boissier, S., et al.\ 2011, \mnras, 415, 1797

\bibitem[Crook et al.(2007)]{2007ApJ...655..790C} Crook, A.~C., Huchra, J.~P., Martimbeau, N., et al.\ 2007, \apj, 655, 790
    
\bibitem[Crook et al.(2008)]{2008ApJ...685.1320C} Crook, A.~C., Huchra, J.~P., Martimbeau, N., et al.\ 2008, \apj, 685, 1320

\bibitem[Duffy et al.(2012)]{2012MNRAS.426.3385D} Duffy, A.~R., Meyer, M.~J., Staveley-Smith, L., et al.\ 2012, \mnras, 426, 3385
    
\bibitem[Eke et al.(2004)]{2004MNRAS.348..866E} Eke, V.~R., Baugh, C.~M., Cole, S., et al.\ 2004, \mnras, 348, 866

\bibitem[Eke et al.(2004)]{2004MNRAS.348..866E} Eke, V.~R., Baugh, C.~M., Cole, S., et al.\ 2004, \mnras, 348, 866

\bibitem[Gavazzi et al.(2013)]{2013A&A...553A..89G} Gavazzi, G., Fumagalli, M., Fossati, M., et al.\ 2013, \aap, 553, A89

\bibitem[Giovanelli et al.(2005)]{2005AJ....130.2598G} Giovanelli, R., Haynes, M.~P., Kent, B.~R., et al.\ 2005, \aj, 130, 2598

\bibitem[Gunn \& Gott(1972)]{1972ApJ...176....1G} Gunn, J.~E. \& Gott, J.~R.\ 1972, \apj, 176, 1

\bibitem[Guo et al.(2015)]{2015MNRAS.453.4368G} Guo, H., Zheng, Z., Zehavi, I., et al.\ 2015, \mnras, 453, 4368

\bibitem[Guo et al.(2017)]{2017ApJ...846...61G} Guo, H., Li, C., Zheng, Z., et al.\ 2017, \apj, 846, 61

\bibitem[Guo et al.(2020)]{2020ApJ...894...92G} Guo, H., Jones, M.~G., Haynes, M.~P., et al.\ 2020, \apj, 894, 92

\bibitem[Haynes et al.(2011)]{2011AJ....142..170H} Haynes, M.~P., Giovanelli, R., Martin, A.~M., et al.\ 2011, \aj, 142, 170

\bibitem[Haynes et al.(2018)]{2018ApJ...861...49H} Haynes, M.~P., Giovanelli, R., Kent, B.~R., et al.\ 2018, \apj, 861, 49

\bibitem[Hess \& Wilcots(2013)]{2013AJ....146..124H} Hess, K.~M. \& Wilcots, E.~M.\ 2013, \aj, 146, 124

\bibitem[Hibbard \& van Gorkom(1996)]{1996AJ....111..655H} Hibbard, J.~E. \& van Gorkom, J.~H.\ 1996, \aj, 111, 655

\bibitem[Huang et al.(2012)]{2012ApJ...756..113H} Huang, S., Haynes, M.~P., Giovanelli, R., et al.\ 2012, \apj, 756, 113

\bibitem[Jing et al.(1998)]{1998ApJ...494....1J} Jing, Y.~P., Mo, H.~J., \& B{\"o}rner, G.\ 1998, \apj, 494, 1

\bibitem[Jones et al.(2018)]{2018MNRAS.477....2J} Jones, M.~G., Haynes, M.~P., Giovanelli, R., et al.\ 2018, \mnras, 477, 2

\bibitem[Kannappan(2004)]{2004ApJ...611L..89K} Kannappan, S.~J.\ 2004, \apjl, 611, L89
   
\bibitem[Katsianis et al.(2017)]{2017MNRAS.472..919K} Katsianis, A., Blanc, G., Lagos, C.~P., et al.\ 2017, \mnras, 472, 919
\bibitem[Katsianis et al.(2017)]{2017MNRAS.464.4977K} Katsianis, A., Tescari, E., Blanc, G., et al.\ 2017, \mnras, 464, 4977

\bibitem[Katsianis et al.(2019)]{2019ApJ...879...11K} Katsianis, A., Zheng, X., Gonzalez, V., et al.\ 2019, \apj, 879, 11

\bibitem[Katsianis et al.(2020)]{2020MNRAS.492.5592K} Katsianis, A., Gonzalez, V., Barrientos, D., et al.\ 2020, \mnras, 492, 5592

\bibitem[Koribalski(2012)]{2012PASA...29..359K} Koribalski, B.~S.\ 2012, \pasa, 29, 359

\bibitem[Lavaux \& Hudson(2011)]{2011MNRAS.416.2840L} Lavaux, G. \& Hudson, M.~J.\ 2011, \mnras, 416, 2840

\bibitem[Li et al.(2012)]{2012MNRAS.424.1471L} Li, C., Kauffmann, G., Fu, J., et al.\ 2012, \mnras, 424, 1471

\bibitem[Lim et al.(2017)]{2017MNRAS.470.2982L} Lim, S.~H., Mo, H.~J., Lu, Y., et al.\ 2017, \mnras, 470, 2982

\bibitem[Lopez et al.(2020)]{2020MNRAS.491.4442L} Lopez, S., Tejos, N., Barrientos, L.~F., et al.\ 2020, \mnras, 491, 4442

\bibitem[Lu et al.(2016)]{2016ApJ...832...39L} Lu, Y., Yang, X., Shi, F., et al.\ 2016, \apj, 832, 39

\bibitem[Martin et al.(2010)]{2010ApJ...723.1359M} Martin, A.~M., Papastergis, E., Giovanelli, R., et al.\ 2010, \apj, 723, 1359

\bibitem[Masui et al.(2013)]{2013ApJ...763L..20M} Masui, K.~W., Switzer, E.~R., Banavar, N., et al.\ 2013, \apjl, 763, L20

\bibitem[McGaugh(2012)]{2012AJ....143...40M} McGaugh, S.~S.\ 2012, \aj, 143, 40

\bibitem[Merritt(1983)]{1983ApJ...264...24M} Merritt, D.\ 1983, \apj, 264, 24

\bibitem[Meyer et al.(2004)]{2004MNRAS.350.1195M} Meyer, M.~J., Zwaan, M.~A., Webster, R.~L., et al.\ 2004, \mnras, 350, 1195

\bibitem[Moore et al.(1996)]{1996Natur.379..613M} Moore, B., Katz, N., Lake, G., et al.\ 1996, \nat, 379, 613

\bibitem[Nan et al.(2011)]{2011IJMPD..20..989N} Nan, R., Li, D., Jin, C., et al.\ 2011, International Journal of Modern Physics D, 20, 989

\bibitem[Papastergis et al.(2011)]{2011ApJ...739...38P} Papastergis, E., Martin, A.~M., Giovanelli, R., et al.\ 2011, \apj, 739, 38
    
\bibitem[Piotrowska et al.(2020)]{2020MNRAS.492L...6P} Piotrowska, J.~M., Bluck, A.~F.~L., Maiolino, R., et al.\ 2020, \mnras, 492, L6

\bibitem[Planck Collaboration et al.(2016)]{2016A&A...594A..27P} Planck Collaboration, Ade, P.~A.~R., Aghanim, N., et al.\ 2016, \aap, 594, A27

\bibitem[Rasmussen et al.(2012)]{2012ApJ...757..122R} Rasmussen, J., Mulchaey, J.~S., Bai, L., et al.\ 2012, \apj, 757, 122
\bibitem[Rasmussen et al.(2012)]{2012ApJ...747...31R} Rasmussen, J., Bai, X.-N., Mulchaey, J.~S., et al.\ 2012, \apj, 747, 31
\bibitem[Saintonge(2007)]{2007AJ....133.2087S} Saintonge, A.\ 2007, \aj, 133, 2087
\bibitem[Saintonge et al.(2017)]{2017ApJS..233...22S} Saintonge, A., Catinella, B., Tacconi, L.~J., et al.\ 2017, \apjs, 233, 22
\bibitem[Serra et al.(2012)]{2012MNRAS.422.1835S} Serra, P., Oosterloo, T., Morganti, R., et al.\ 2012, \mnras, 422, 1835
\bibitem[Solanes et al.(2001)]{2001ApJ...548...97S} Solanes, J.~M., Manrique, A., Garc{\'\i}a-G{\'o}mez, C., et al.\ 2001, \apj, 548, 97
\bibitem[Stark et al.(2016)]{2016ApJ...832..126S} Stark, D.~V., Kannappan, S.~J., Eckert, K.~D., et al.\ 2016, \apj, 832, 126
\bibitem[Strateva et al.(2001)]{2001AJ....122.1861S} Strateva, I., Ivezi{\'c}, {\v{Z}}., Knapp, G.~R., et al.\ 2001, \aj, 122, 1861
\bibitem[Switzer et al.(2013)]{2013MNRAS.434L..46S} Switzer, E.~R., Masui, K.~W., Bandura, K., et al.\ 2013, \mnras, 434, L46
\bibitem[Tacconi et al.(2018)]{2018ApJ...853..179T} Tacconi, L.~J., Genzel, R., Saintonge, A., et al.\ 2018, \apj, 853, 179
\bibitem[Taylor et al.(2012)]{2012MNRAS.423..787T} Taylor, R., Davies, J.~I., Auld, R., et al.\ 2012, \mnras, 423, 787
\bibitem[Tinker et al.(2008)]{2008ApJ...688..709T} Tinker, J., Kravtsov, A.~V., Klypin, A., et al.\ 2008, \apj, 688, 709
\bibitem[Tr{\v{c}}ka et al.(2020)]{2020MNRAS.494.2823T} Tr{\v{c}}ka, A., Baes, M., Camps, P., et al.\ 2020, \mnras, 494, 2823
\bibitem[Vale \& Ostriker(2004)]{2004MNRAS.353..189V} Vale, A. \& Ostriker, J.~P.\ 2004, \mnras, 353, 189
\bibitem[Wang et al.(2014)]{2014MNRAS.439..611W} Wang, L., Yang, X., Shen, S., et al.\ 2014, \mnras, 439, 611
\bibitem[Wang et al.(2015)]{2015MNRAS.449.2010W} Wang, E., Wang, J., Kauffmann, G., et al.\ 2015, \mnras, 449, 2010
\bibitem[Wechsler \& Tinker(2018)]{2018ARA&A..56..435W} Wechsler, R.~H. \& Tinker, J.~L.\ 2018, \araa, 56, 435
\bibitem[Yan et al.(2003)]{2003ApJ...598..848Y} Yan, R., Madgwick, D.~S., \& White, M.\ 2003, \apj, 598, 848
\bibitem[Yang et al.(2003)]{2003MNRAS.339.1057Y} Yang, X., Mo, H.~J., \& van den Bosch, F.~C.\ 2003, \mnras, 339, 1057
\bibitem[Yang et al.(2005)]{2005MNRAS.356.1293Y} Yang, X., Mo, H.~J., van den Bosch, F.~C., et al.\ 2005, \mnras, 356, 1293
    
\bibitem[Yang et al.(2006)]{2006MNRAS.369.1293Y} Yang, X., van den Bosch, F.~C., Mo, H.~J., et al.\ 2006, \mnras, 369, 1293
\bibitem[Yang et al.(2007)]{2007ApJ...671..153Y} Yang, X., Mo, H.~J., van den Bosch, F.~C., et al.\ 2007, \apj, 671, 153
\bibitem[Yang et al.(2008)]{2008ApJ...676..248Y} Yang, X., Mo, H.~J., \& van den Bosch, F.~C.\ 2008, \apj, 676, 248
\bibitem[Yang et al.(2009)]{2009ApJ...695..900Y} Yang, X., Mo, H.~J., \& van den Bosch, F.~C.\ 2009, \apj, 695, 900
\bibitem[Yang et al.(2012)]{2012ApJ...752...41Y} Yang, X., Mo, H.~J., van den Bosch, F.~C., et al.\ 2012, \apj, 752, 41
\bibitem[Yoon \& Rosenberg(2015)]{2015ApJ...812....4Y} Yoon, I. \& Rosenberg, J.~L.\ 2015, \apj, 812, 4
\bibitem[York et al.(2000)]{2000AJ....120.1579Y} York, D.~G., Adelman, J., Anderson, J.~E., et al.\ 2000, \aj, 120, 1579
\bibitem[Zehavi et al.(2011)]{2011ApJ...736...59Z} Zehavi, I., Zheng, Z., Weinberg, D.~H., et al.\ 2011, \apj, 736, 59
\bibitem[Zhang et al.(2009)]{2009MNRAS.397.1243Z} Zhang, W., Li, C., Kauffmann, G., et al.\ 2009, \mnras, 397, 1243
\bibitem[Zhang et al.(2007)]{2007ApJ...655..851Z} Zhang, W., Kong, X., Li, C., et al.\ 2007, \apj, 655, 851
\bibitem[Zhang et al.(2020)]{2020ApJ...888...33Z} Zhang, H., Yang, X., Zaritsky, D., et al.\ 2020, \apj, 888, 33
\bibitem[Zheng et al.(2005)]{2005ApJ...633..791Z} Zheng, Z., Berlind, A.~A., Weinberg, D.~H., et al.\ 2005, \apj, 633, 791
\bibitem[Zwaan et al.(2005)]{2005MNRAS.359L..30Z} Zwaan, M.~A., Meyer, M.~J., Staveley-Smith, L., et al.\ 2005, \mnras, 359, L30
\bibitem[van den Bosch et al.(2005)]{2005MNRAS.356.1233V} van den Bosch, F.~C., Yang, X., Mo, H.~J., et al.\ 2005, \mnras, 356, 1233


\end{thebibliography}

\label{lastpage}

\end{document}